\documentclass[reprint, amsmath,amssymb,aps]{revtex4-1}
\usepackage{graphicx}
\usepackage{dcolumn}
\usepackage{bm}
\usepackage{hyperref}
\newcommand{\qm}[1]{``#1''}

\begin{document}

\preprint{APS/123-QED}

\title{Lagrangian formulation of the general relativistic Poynting-Robertson effect}

\author{Vittorio De Falco$^{1,2}$, Emmanuele Battista$^{3,4}$, and Maurizio Falanga$^{1,2}$.}

\affiliation{$^1$ International Space Science Institute, Hallerstrasse 6, 3012 Bern, Switzerland.
\email{deltafi.mat@gmail.com}\\
$^2$ Departement Physik, Universit\"at Basel, Klingelbergstrasse 82, 4056 Basel, Switzerland\\
$^3$ Universit\'a degli studi di Napoli \qm{Federico II}, Dipartimento di Fisica \qm{Ettore Pancini}, Complesso Universitario di Monte S. Angelo, Via Cintia Edificio 6, 80126 Napoli, Italy\\
$^4$ Istituto Nazionale di Fisica Nucleare, Sezione di Napoli, Complesso Universitario di Monte S. Angelo,
Via Cintia Edificio 6, 80126 Napoli, Italy\\}

\date{\today}

\begin{abstract}
We propose the Lagrangian formulation for describing the motion of a test particle in a general relativistic, stationary, and axially symmetric spacetime. The test particle is also affected by a radiation field, modeled as a coherent flux of photons traveling along the null geodesics of the background spacetime, including the general relativistic Poynting-Robertson effect. The innovative part of this work is to prove the existence of the potential linked to the dissipative action caused by the Poynting-Robertson effect in General Relativity through the help of an integrating factor, depending on the energy of the system. Generally such kinds of inverse problems involving dissipative effects might not admit a Lagrangian formulation, especially in General Relativity there are no examples of such attempts in literature so far. We reduce this general relativistic Lagrangian formulation to the classic case in the weak field limit. This approach facilitates further studies in improving the treatment of the radiation field and it contains for example some implications for a deeper comprehension of the gravitational waves.
\end{abstract}
                              
\maketitle
\section{Introduction}
\label{sec:intro}
In high-energy astrophysics, it is important to examine the effects of the radiation field on the motion of the matter surrounding a compact object. Such radiation field can be arisen from: a boundary layer around a neutron star \citep[NS;][]{Inogamov1999}, a thermonuclear flash (type-I X-ray burst) occurring on an accreting NS surface \citep{Lewin1993}, or a hot corona around a black hole (BH) in X-ray binary systems \citep{Fabian2015}. Many observations confirm that this kind of radiation field, beside to exert an outward radial force, opposite to the gravitational pull from the compact object, can generate also a drag force, produced during the process of absorption and re-emission of the radiation from the affected body \citep{Ballantyne2004,Ballantyne2005,Worpel2013,Keek2014,Ji2014,Worpel2015}. This force plays an important role in removing angular momentum and energy from the considered body, thus altering its motion. This phenomenon, also known as Poynting-Robertson (PR) effect, was first investigated in classical Newtonian gravity by \citet{Poynting1903}, extended then to Special Relativity by \citet{Robertson1937}, and finally generalized to General Relativity (GR) for a general stationary axially symmetric spacetime by \citet{Bini2009,Bini2011}. The results of all these calculations show, that the radiation alters substantially the velocity of the matter, even for luminosities lower than the Eddington limit.

The PR effect has been applied to clarify some puzzling observations in the accretion physics. Among the works on this subject, it is fundamental to mention: the studies of \citet{Abramowicz1990} to model relativistic superluminal jets and NS winds in the exterior Schwarzschild metric; the model, developed by \citet{Walker1989,Walker1992}, to investigate the way in which a type I X-ray burst, on the surface of an accreting NS, induces an increased mass inflow rate in the inner edge of the accretion disc; the anlyses of \citet{Lamb1995,Miller1996} to understand how the radiation field and the PR effect influence the velocity field of the accreting matter around a slowly rotating compact star and how it can be related to the star spin evolution; the subsequent calculations of \citet{Miller1998}, who determined the inner disc radius at which the radiation effects cause the radial velocity to exceed the sound speed; the recent model proposed by Bakala et al. (2018, A\&A submitted), who exploit the general relativistic description of the PR effect, combined with respect to the standard accretion disc theory, to build up a model able to follow the dynamical evolution of an accretion disc around a compact object when it is affected by a constant luminosity.

The PR effect theory has never been treated from a Lagrangian point of view at the best of our knowledge. This approach is advantageous for the following reasons: ($i$) it is a general, elegant, and effective methodology to attain at the \emph{pure} equations of motion (EoMs), namely once the kinetic, $\mathbb{T}$, and potential, $\mathbb{V}$, energies of the system are given, the constraint forces and the generalized forces, $Q_h$, are identified, and the generalized coordinates, $(q_h,\dot{q}_h)$, are chosen, we can analytically calculate the EoMs through the \emph{Euler-Lagrange equations} (ELEs), i.e.:
\begin{equation} \label{eq:ELeq}
\frac{d}{dt}\frac{\partial\mathcal{L}}{\partial \dot{q}_h}-\frac{\partial\mathcal{L}}{\partial q_h}=Q_h,
\end{equation}
where $\mathcal{L}=\mathbb{T}-\mathbb{V}$ is the \emph{Lagrangian function}; ($ii$) it places emphasis on the geometrical structure of the problem; ($iii$) the formulation of the motion is expressed in integral terms, not anymore in differential form \citep{Chow1995,Goldstein2002}. The nature of the ELEs, Eq. (\ref{eq:ELeq}), admits two points of view, depending whether one consider the \emph{direct or inverse problem}. The direct problem coincides with the ordinary approach, where given the Lagrangian function the EoMs are obtained computing the ELEs, seen as second-order ordinary differential equations \citep[ODEs; ][]{Chow1995,Goldstein2002}; on the contrary, the inverse problem consists in determining a possible Lagrangian, that gives, through the ELEs, the assigned EoMs. Now, the ELEs become a set of second-order partial differential equations \citep[PDEs;][]{Santilli1978,Santilli1979,Lopuszanski1999}. 

Regarding the forces acting on a system, $Q_h$, they can be classified in two types: \emph{conservative} and \emph{non-conservative}. The conservative forces can be written as the generalized coordinates gradient of a potential, $Q_h=-\partial \mathbb{V}/\partial q_h$, and can be englobed in the potential energy, $\mathbb{V}$, of the Lagrangian function, $\mathcal{L}$, e.g., all the central force fields \citep{Chow1995,Goldstein2002}. The non-conservative forces can be divided ulteriorly in two subclasses: \emph{generalized forces} and \emph{dissipative forces}. The first ones are forces deriving from a generalized potential function, $\mathbb{V}$, such that $\partial\mathbb{V}/\partial q_h-d/dt(\partial\mathbb{V}/\partial \dot{q}_h)=Q_h$, and they can be included in the potential energy, $\mathbb{V}$, of the Lagrangian function, $\mathcal{L}$  \citep{Chow1995,Goldstein2002}. This potential finds its applications in the Lorentz force for an electromagnetic field and in general for all gyroscopic-like forces, or \emph{divergence-free fields}. Instead, the second forces do not admit a potential function, but they can be written as the velocity gradient of a \emph{Reyleigh dissipative function}, $\mathcal{F}$, i.e., $Q_h=-\partial \mathcal{F}/\partial \dot{q}_h$ \citep{Chow1995,Goldstein2002,Minguzzi2015}. Therefore, in this case we have to assign the Lagrangian and the Reyleigh dissipative potential function, expressing how the dissipative forces act on the system. However, there exists forces not collocabile in any of the groups mentioned above, since they might admit a potential function by adding an \emph{integrating factor}. For example, in classical thermodynamics where the increment of heat multiplied by the inverse of the temperature gives an exact differential form represented physically by the entropy \citep{Courant1962,Santilli1979}. 

The aim of this paper is to develop the Lagrangian formalism of the general relativistic PR effect \citep{Bini2009,Bini2011}, which requires the use of different techniques. The paper is structured as follows: in Sect. \ref{sec:classicPR}, we present the Lagrangian formulation of the classical PR effect; in Sect. \ref{sec:GR_formalism} we recall the relativity of observer splitting formalism, useful to derive the general relativistic PR EoMs; then in Sect. \ref{sec:GRPR} we derive its Lagrangian formulation; the conclusions are drawn in Sect. \ref{sec:conclusions}. 

\section{The classical Poynting-Robertson effect}
\label{sec:classicPR}
The classical radiation drag force was described and introduced by \citet{Poynting1903} and \citet{Robertson1937}. Such phenomenon considers a small spherical test particle of radius $a$, density $\rho$, and with a fully absorbing surface. This is sufficiently small in size, such that the temperature is homogeneous and uniform through all the body, and adequately greater than the wavelength dimensions, otherwise the body cannot absorb the radiation. This test particle moves around the Sun affected by the gravitational pull, the radiation force, and the radiation drag force. The related test particle EoMs, written in polar coordinates, $(r,\varphi)$, are \citep[see][and Eqs. (3.5) in \citet{Robertson1937} for the right correction factor in the radiation drag force term]{Poynting1903}:
\begin{eqnarray} 
&&\ddot{r}-r\dot{\varphi}^2+\frac{GM-Ac}{r^2}=-2A\frac{\dot{r}}{r^2}, \label{eqm1}\\
&&r^2\dot{\varphi}=L_0-A\varphi, \label{eqm2} 
\end{eqnarray}
where the dot means the derivative with respect to time, $G$ is the gravitational constant, $M$ the mass of the Sun, $c$ the speed of the light in the vacuum, $L_0\equiv[r(t)^2\dot{\varphi}(t)]_{t=t_0}$ is a constant of integration representing the specific angular momentum at the initial time, and $A=Sb^2/(6c^2\rho a)$ is the repulsion contribution of the radiation force in $-Ac/r^2$ and the specific angular momentum removed from the test particle due to PR effect in $-2A\dot{r}/r^2$, with $S$ being the energy emitted by the Sun per area and time, and $b$ the distance Sun--Earth. Naturally, this treatment can be easily generalised around other massive radiating sources, choosing thus the relative parameters accordingly. We determine the Lagrangian function $\mathcal{L}_{\rm C}=\mathcal{L}_{\rm C}(q_h,\dot{q}_h)$, depending on the lagrangian coordinates $q_1=r$ and $q_2=\varphi$ and $\dot{q}_h=d q_h/dt$, in presence of the forces $Q_h$, where $Q_1=-2A\dot{r}/r^2$ and $Q_2=-A\dot{\varphi}$, such that the respective ELEs will give the test particle EoMs, Eqs. (\ref{eqm1}) -- (\ref{eqm2}). We note that the forces $Q_h$ are dissipative, because mathematically they depend on the velocity field and physically the energy is removed from the particle when the motion takes place. This force can be written in terms of the Rayleigh dissipative function, $\mathcal{F}_{\rm C}$. Therefore, the Lagrangian function is constituted by the kinetic, $\mathbb{T}_{\rm C}$, and potential, $\mathbb{V}_{\rm C}$, components:
\begin{equation} \label{eq:lagrangian}
\mathcal{L}_{\rm C}\equiv \mathbb{T}_{\rm C}-\mathbb{V}_{\rm C}=\frac{1}{2}\left(\dot{r}^2+r^2\dot{\varphi}^2\right)+\frac{GM-A}{r}.
\end{equation}
Instead, the Reyleigh dissipative function, $\mathcal{F}_{\rm C}$, can be determined by verifying whether the radiation differential form, whose components derived from Eqs. (\ref{eqm1}) -- (\ref{eqm2}), are given by: 
\begin{equation} \label{eq:pot_comp1}
\frac{\partial \mathcal{F}_{\rm C}}{\partial \dot{r}}=-\frac{A\dot{r}}{r^2}, \qquad \frac{\partial \mathcal{F}_{\rm C}}{\partial \dot{\varphi}}=-A\dot{\varphi},
\end{equation} 
is close and the domain, where it is defined, is simply connected. Such differential form is defined on all the space $\mathbb{R}^2$ minus the origin, where it is located the Sun. This domain results to be simply connected, because, in polar coordinates, it transforms into a rectangle, defined by $r\in(0,+\infty]$ and $\varphi\in[0,2\pi]$. In addition, it is immediate to verify that it is a close form, i.e., the cross derivatives are equal. Therefore, such potential is obtained by integrating Eqs. (\ref{eq:pot_comp1}), constituting a decoupled system of PDEs.
$\mathcal{F}_{\rm C}$ is easily found:
\begin{equation} \label{eq:pot_comp2}
\mathcal{F}_{\rm C}(\dot{r},\dot{\varphi})=\frac{A}{2}\left(\frac{\dot{r}^2}{r^2}+\dot{\varphi}^2\right)+const,
\end{equation} 
as an homogeneous function of order two in $(\dot{r},\dot{\varphi})$, completely determined once the initial conditions are set. 

\section{The relativistic Poynting-Robertson effect}
\label{sec:GR_formalism}
In classical mechanics the centrifugal forces are conceived to be fictitious inertial forces, that manifest themselves all the time we are observing the dynamics of an object in a rotating reference frame. Inertial forces have been topics of great interests in GR, because: ($i$) there is a close similarity between the gravitational forces, experienced locally on a massive body, and the fictitious pseudo-forces, felt by an observer in a non-inertial accelerated reference frame (argument based on the \emph{equivalence principle}); ($ii$) there is a strong analogy between the gravitational forces and the electromagnetism description, the so-called \emph{gravitoelectromagnetism} \citep[][]{DeFelice1971}. Therefore, there have been many efforts to generalize the concept of centrifugal force to stationary \citep{Abramowicz1988,Abramowicz1990,DeFelice1991a}, axially symmetric \citep{Prasanna1990,DeFelice1991b,Iyer1993,DeFelice1995,Barrabes1995} and finally also to arbitrary spacetimes \citep{Abramowicz1993}, encountering, however, several difficulties. The flaw of such attempts reside in the employment of the \emph{direct spacetime approach}, where the interpretation of the dynamical variables depends on further quantities \citep[see e.g.,][]{Abramowicz1988,Abramowicz1990,Abramowicz1993}. 

The successful approach, in terms of comprehensibility and clearness about the outcoming results, revealed to be the \emph{relativity of observer splitting formalism}, based on the full orthogonal splitting of the test particle motion relative to the observer in: ($i$) \qm{4=3+1}: local rest space and local time direction; ($ii$) \qm{{\it3=2+1}}: transverse and longitudinal components of the local rest space \citep[see e.g.,][]{Bini1997a,Bini1998}. Such formalism entails several advantages: it relies on a logical and unambiguous mathematical structure, it offers a natural link with respect to the classical case and, in the same time, provides an explicit physical interpretation of the involved terms \citep[see e.g.,][]{Bini1997b,Bini1998}. In addition, the general relativistic description, at the contrary of the classical framework, mixes the gravitational field, due to the presence of matter, with those of the accelerated motion of the observers. Therefore, it is significative to choose a Frenet-Serret frame, where it is possible to coherently split the different contributions and make sense to the splitting, reproducing thus the classical case \citep{Bini1997a,Bini1999}. 

In the following sections, we present for completeness the modern approach to derive the general relativistic PR EoMs, focussing the attention on its geometrical aspects. We show how this general relativistic formalism, to describe the non-inertial relative motions, recovers the line of though of the classical formalism. 

\subsection{Classical formalism and non-inertial relative motions}
We consider two reference frames $\mathbb{R}\equiv\left\{O,x,y,z\right\}$ and $\mathbb{R'}\equiv\left\{O',x',y',z'\right\}$ in relative motion to each other, observing the dynamical trajectory described by a point $P$. We call $\mathbf{r}(t)=P(t)-O$ and $\mathbf{r'}(t)=P(t)-O'$ the radius vectors in the reference frames $\mathbb{R}$ and $\mathbb{R'}$, respectively. Thus, we have the following relationship between the positions $\mathbf{r}=\mathbf{r'}+OO'$. Using the Poisson's formula on the versors of the reference frame $\mathbb{R'}$, i.e., $d\mathbf{u}/dt=\pmb{\omega}\times\mathbf{u}$, where $\mathbf{u}$ is a versor and $\pmb{\omega}$ is the angular velocity associated to the variation of $\mathbf{u}$, we have the relation between the velocities:
\begin{equation} \label{eq:velocity}
\mathbf{v}=\mathbf{v_{O'}}+\mathbf{v'}+\pmb{\omega}\times\mathbf{r'},
\end{equation}
where $\mathbf{v_t}=\mathbf{v_{O'}}+\mathbf{v'}$ is the translatory velocity and $\pmb{\omega}\times\mathbf{r'}$ is the rotating velocity. Now passing to the accelerations, we have:
\begin{equation} \label{eq:acceleration}
\mathbf{a}=\mathbf{a_{O'}}+\mathbf{a'}+\pmb{\omega}\times\pmb{\omega}\times\mathbf{r'}+2\pmb{\omega}\times\mathbf{v'},
\end{equation}
where $\mathbf{a_t}=\mathbf{a_{O'}}+\mathbf{a'}$ is the translatory acceleration, $\pmb{\omega}\times\pmb{\omega}\times\mathbf{r'}$ is the centrifugal force, and $2\pmb{\omega}\times\mathbf{v'}$ is the Coriolis force. Eqs. (\ref{eq:velocity}) -- (\ref{eq:acceleration}) are well known in the literature, where all the components have a precise and clear physical meaning \citep{Sommerfeld1964,Arnold2013}. 

\subsection{General relativistic formalism}
\label{sec:observer}
We consider a Riemannian manifold endowed with a Lorentzian metric, $g_{\alpha\beta}$, with signature $(-,+,+,+)$, a symmetric Levi-Civita connection, $\Gamma^\alpha_{\beta\gamma}$, and a covariant derivative, $\nabla_\alpha$. We also consider a family of observers with a four-velocity, defined by a future-pointing unit timelike vector field $u^\alpha$ ($u_\alpha u^\alpha=-1$). Its proper time, $\tau_u$, parametrizes the world lines, that are integral curves of $u^\alpha$, the so-called \emph{observer congruence} \citep[see Chap. 6 of][]{Stephani2003}. 
 
\subsubsection{3+1 splitting}
In order to orthogonally decompose each tangent space into local rest space and local time in the direction of the observer $u^\alpha$, we project all the quantities in the hypersurface orthogonal to $u^\alpha$, through the projector operator $P(u)_{\alpha\beta}=g_{\alpha\beta}+u_\alpha u_\beta$. All the tensors having no components along $u^\alpha$, are termed \emph{spatial}. 

\subsubsection{Kinematical decomposition of observer congruence}
We define the acceleration vector related to the observer $u^\alpha$, given by $a(u)^\alpha=u^\beta \nabla_\beta u^\alpha$. Using $u_\alpha u^\alpha=-1$, it can be easily proved, that this acceleration has the propriety to be orthogonal to its velocity field $u^\alpha$, i.e., $u^\alpha a(u)_\alpha=0$ \citep{Misner1973}. We note that $(a(u)^\alpha u_\beta+\nabla_\beta u^\alpha)u^\beta=0$, therefore the term in parenthesis is spatial and can be decomposed into its symmetric and antisymmetric parts \citep{Wald1984,Stephani2003}:
\begin{equation}
\nabla_\alpha u_\beta=-a(u)_\alpha u_\beta+\theta_{\alpha\beta}+\omega_{\alpha\beta},
\end{equation}
where $\theta_{\alpha\beta}=\nabla_{(\beta}u_{\alpha)}a_{(\alpha}u_{\beta)}\equiv P^\mu_\alpha P^\nu_\beta \nabla_{(\nu} u_{\mu)}$ is the expansion tensor, describing how an initial spherical cloud of test particles becomes distorted into an ellipsoidal shape; $\omega_{\alpha\beta}=\nabla_{[\beta}u_{\alpha]}a_{[\alpha}u_{\beta]}\equiv P^\mu_\alpha P^\nu_\beta \nabla_{[\nu} u_{\mu]}$ is the vorticity tensor, representing how an initial spherical cloud of test particles tends to rotate. We have used the following notations: $(A,B)=\frac{1}{2}(AB+BA)$ and $[A,B]=\frac{1}{2}(AB-BA)$. This is the so-called \emph{kinematical decomposition of the observer congruence}.

\subsubsection{Spatial derivative operators}
\begin{figure} [h!]
\includegraphics[scale=0.39]{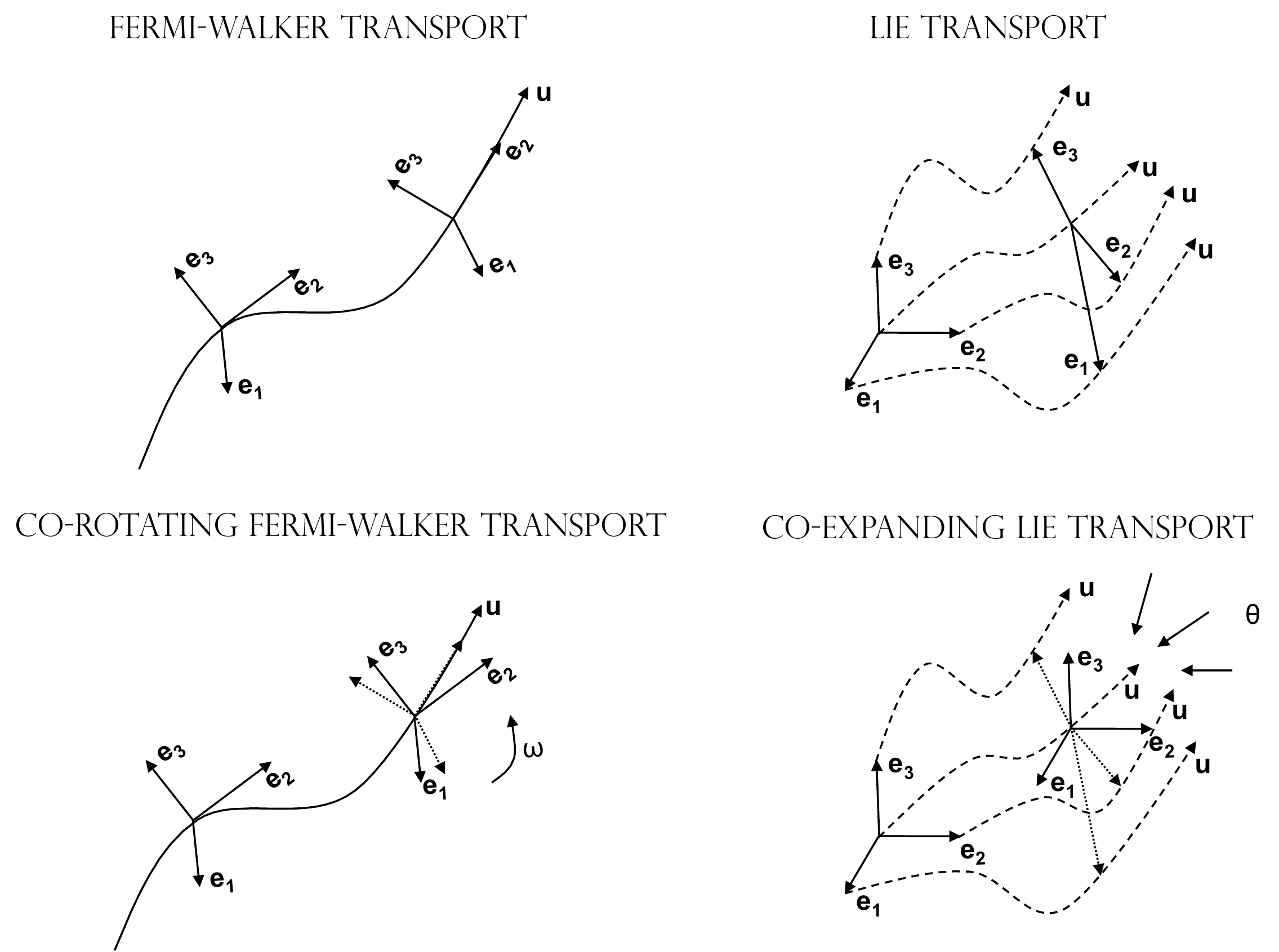}
\caption{Geometrical representation of the Fermi-Walker, Lie, co-rotating Fermi-Walker, and co-expanding Lie transport.}
\label{fig:SDO}
\end{figure}

It is now appropriate to introduce a set of spatial derivative operators, through the projector $P(u)_{\alpha\beta}$, in order to achieve the 2+1 splitting. Given any spatial vector field, $X^\alpha$, and a generic vector, $v^\alpha$, we define \citep{Bini1997a,Bini1998,Bini1999,Bini2010} 
\begin{itemize}
\item the spatial Lie derivative: 
\begin{equation}
\begin{aligned}
\mathcal{L}(u)_Xv^\alpha&=P(u)\mathcal{L}_Xv^\alpha\\
&=P(u)^\alpha_\gamma \left( X^\beta\nabla_\beta v^\gamma+v^\beta\nabla_\beta X^\gamma \right);
\end{aligned}
\end{equation}
\item the spatial covariant derivative: 
\begin{equation}
\nabla(u)_\beta v^\alpha=P(u)^\alpha_\delta P(u)_\beta^\gamma\nabla_\gamma v^\delta;
\end{equation}
\item the temporal Lie derivative: 
\begin{equation}
\nabla_{\rm (Lie)}(u)v^\alpha=P(u)\mathcal{L}_u v^\alpha;
\end{equation}
\item the temporal Fermi-Walker derivative: 
\begin{equation}
\nabla_{\rm (fw)}(u)v^\alpha=P(u)^\alpha_\gamma u^\beta\nabla_\beta v^\gamma.
\end{equation}
\end{itemize}
In the definitions reported above, we have introduced two ways to perform the derivatives, i.e., the \emph{Fermi-Walker} and \emph{Lie transport}. The Fermi-Walker transport with respect to the observer congruence $u^\alpha$ moves rigidly a spatial tetrad, $(\mathbf{e_1},\mathbf{e_2},\mathbf{e_3})$, along the world line described by the vector $u^\alpha$ \citep[see Fig. \ref{fig:SDO} and][]{Misner1973}; instead the Lie transport with respect to the observer congruence $u^\alpha$ evolves a spatial tetrad, $(\mathbf{e_1},\mathbf{e_2},\mathbf{e_3})$, following the geodesic flux described by $u^\alpha$ \citep[see Fig. \ref{fig:SDO} and][]{Hawking1973,Stephani2003}. 
There is another transport obtained combining the Fermi-Walker (or Lie) transport with respect to the kinematical decompositions of the observer, i.e., the temporal \emph{co-rotating Fermi-Walker} (or the \emph{co-expanding Lie}) \emph{derivative} \citep[see Fig. \ref{fig:SDO} and][]{Bini1997a,Bini1998,Bini1999,Bini2010}: 
\begin{equation}
\begin{aligned}
\nabla_{\rm (cfw)}(u)v^\alpha&=\nabla_{\rm(fw)}(u)v^\alpha+\omega(u)^\alpha_\beta v^\beta\\
&=\nabla_{\rm(Lie)}(u)v^\alpha+\theta_\beta^\alpha v^\beta.
\end{aligned}
\end{equation}
This kind of transport with respect to the observer congruence $u^\alpha$ let a spatial tetrad, $(\mathbf{e_1},\mathbf{e_2},\mathbf{e_3})$, unchanged during the evolution, either rigidly along the observer world line and after applying an opportune rotation, or according to the geodesic flow described by $u^\alpha$ and after applying an opportune expansion.

\subsubsection{Nonlinear reference frame}
\label{sec:nrf}
\begin{figure} [h!]
\includegraphics[scale=0.37]{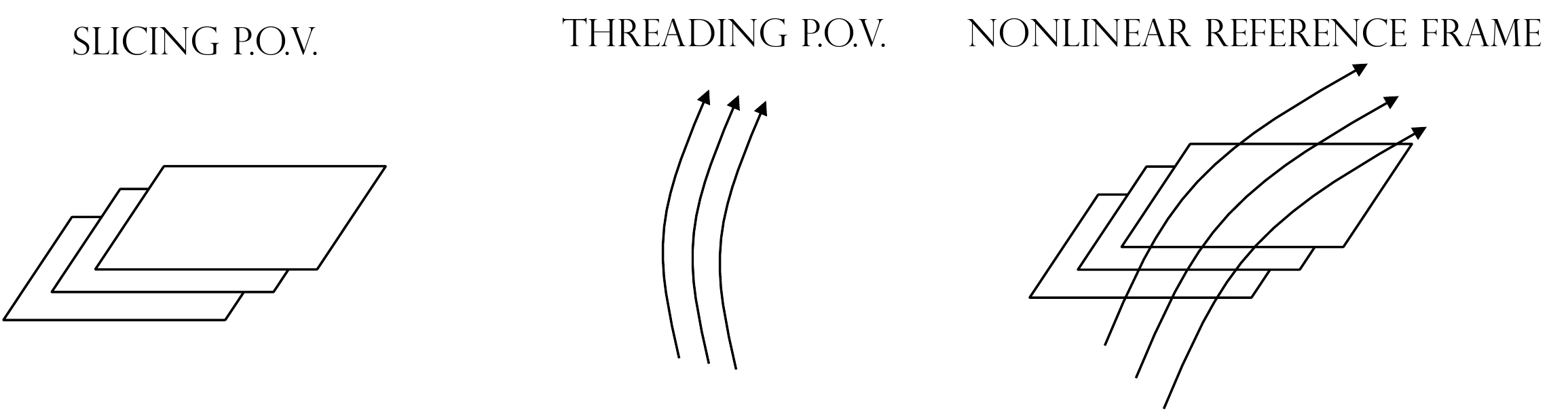}
\caption{Geometrical representation of the slicing and threading point of view, and nonlinear reference frame.}
\label{fig:pov}
\end{figure}

To further split the local rest space we have to adopt an adequate point of view. A full splitting of spacetime manifold requires both a \emph{slicing} of the spacetime in spatial hypersurfaces and a \emph{threading} of the spacetime along the observer congruence. A slicing together with a transversal threading form a structure dubbed \emph{nonlinear reference frame} \citep[see Fig. \ref{fig:pov} and][]{Jantzen1992,Bini1997a,Bini1998}. 
Therefore, we have to introduce a system of coordinates adapted to the observer congruence. Let $\left\{E_a^\alpha\right\}\equiv \left\{E_1^\alpha, E_2^\alpha, E_3^\alpha\right\}$ be a spatial frame, i.e., a basis of each local rest space with respect to the observer congruence $u^\alpha$. The Latin-index refers to the spatial frame components, instead the Greek-index refers to the spacetime components. The frame derivatives of a function, $f$, are denoted by the comma notation:
$$
u(f)=f_{,0},\qquad E_a^\alpha\partial_\alpha f=f_{,\alpha}.
$$  
To perform the derivatives of tensor fields, we define temporal and spatial derivatives of the spatial frame vectors, as well as their Lie brackets \citep[see Eqs. (11.2) in][]{Bini1997a}:
\begin{eqnarray}
&&\nabla_{\rm (tem)}(u)\,E_a^\alpha=C_{\rm (tem)}(u)^b_a\, E_b^\alpha,\ _{\rm tem=fw,\ Lie,\ cfw,}\label{eq:cc1}\\
&&\nabla_{E_a}\,E_b^\alpha=\Gamma(u)^c_{ab}\, E_c^\alpha,\label{eq:cc2}\\
&&\left(P(u)\,[E_a,E_b]\right)^\alpha=C(u)^c_{ab}\, E_c^\alpha,\label{eq:cc3}
\end{eqnarray}
where the following terms may be called: $C_{\rm (tem)}(u)^b_a$ the \qm{temporal constant structures}, $\Gamma(u)^c_{ab}$ the \qm{spatial connections}, and $C(u)^c_{ab}$ the \qm{spatial constant structures}. 

\subsection{Test particle motion}
We consider the motion of a test particle with respect to unit timilike four-velocity, $U^\alpha$, and we describe such motion relative to the observer $u^\alpha$. The world line of such particle is parametrized by the particle's proper time, $\tau_U$, related to the observer proper time, $\tau_{(U,u)}$, by $d\tau_{(U,u)}/d\tau_U=\gamma(U,u)$. Therefore, the four-velocity $U^\alpha$ can be decomposed in the component along $u^\alpha$, $U^{(||_u)}$, and in the spatial components with respect to $u^\alpha$, $[P(u)U]^\alpha$, as:
\begin{equation} \label{eq:U_split}
\begin{aligned}
U(\tau_U)^\alpha&=U(\tau_U)^{(||_u)}\,u^\alpha+[P(u)\,U(\tau_U)]^\alpha\\
&=\gamma(U,u)\,[u^\alpha+\nu(U,u)^\alpha]\\
&=E(U,u)\,u^\alpha+p(U,u)^\alpha,
\end{aligned}
\end{equation}
where $\nu^\alpha\equiv\nu^\alpha(U,u)$ is the relative spatial velocity of $U^\alpha$, $\gamma\equiv\gamma(U,u)=(1-\nu^2)^{-1/2}$ is the Lorentz factor with $\nu\equiv\nu(U,u)=\sqrt{\nu_\alpha \nu^\alpha}$ the module of the relative spatial velocity of $U^\alpha$, $E\equiv E(U,u)=\gamma$ is the energy per unit mass as seen by the observer $u^\alpha$, and $p^\alpha\equiv p(U,u)^\alpha=\gamma(U,u)\nu(U,u)^\alpha$ is the relative spatial momentum of the test particle per unit mass, and $p\equiv p(U,u)=\sqrt{p_\alpha p^\alpha}$ is the module of the relative spatial momentum. 

\subsection{Intrinsic derivative along the test particle curve}
In order to operate along the test particle curve, we define the \emph{intrinsic} or \emph{absolute derivative} of a spatial vector field, $X^\alpha$, along the test particle trajectory simply restricting the action of the covariant derivative on the vector field $X^\alpha$ along the test particle curve, i.e. \citep{Bini1997a,Bini1998,Bini2010}:
\begin{equation}
\frac{DX^\alpha(\tau_U)}{d\tau_U}=\frac{dX^\alpha(\tau_U)}{d\tau_U}+\Gamma^\alpha_{\beta\gamma}\,U(\tau_U)^\beta\, X^\gamma(\tau_U).
\end{equation}
Therefore, we can extend the notions of Fermi-Walker, Lie, and co-rotating Fermi-Walker transport along the test particle curve in the following way \citep{Bini1997a,Bini1998,Bini2010}:
\begin{equation} \label{eq:Dcurve}
\begin{aligned}
&\frac{D_{\rm(tem)}(U,u)\ X^\alpha(\tau_U)}{d\tau_{(U,u)}}=\left[\nabla_{\rm(tem)}(u)+\nu^\beta\,\nabla(u)_\beta\right]X^\alpha(\tau_U)\\ 
&\hspace{3cm}_{\rm tem=fw,\ Lie,\ cfw}.
\end{aligned}
\end{equation}
We note that a factor $\gamma$ is missing in Eq. (\ref{eq:Dcurve}), because it is included into the reparametrization of the particle world line through the formula $d\tau_{(U,u)}/d\tau_U=\gamma(U,u)$.

\subsection{2+1 splitting}
We consider a system of coordinates $\left\{u^\alpha, E^\alpha_a\right\}$ adapted to the observer congruence in a nonlinear reference frame (see Sect. \ref{sec:nrf}). The spatial projection of the test particle's four-acceleration, $a(U)^\alpha=DU^\alpha/d\tau_U$, measured by the the observer congruence, $A(U,u)^\alpha$, is given by $A(U,u)^\alpha=1/\gamma P(u)^\alpha_\beta a(U)^\beta$. Therefore, we have \citep[see Sect. 9 in][]{Bini1997a,Bini1998}:
\begin{equation} \label{eq:S+T}
\begin{aligned}
a(U)^\alpha&=\gamma\, P(U)^\alpha_\beta\, A(U,u)^\beta\\
&=\gamma\, P(U)^\alpha_\beta\,\left\{\frac{D_{\rm (tem)}(U,u)}{d\tau_{(U,u)}}\left[\gamma\, u^\alpha+p(u,U)^\alpha\right]\right\}\\
&=\gamma^2\frac{D_{\rm (tem)}(U,u)}{d\tau_{(U,u)}}u^\alpha+\gamma\frac{D_{\rm (tem)}(U,u)}{d\tau_{(U,u)}}p(u,U)^\alpha.
\end{aligned}
\end{equation}
It is important to note that the term connected with the derivative of the factor $\gamma$ does not appear in Eq. (\ref{eq:S+T}) because it carries a term $u^\alpha$ that it is vanished by the projector, since we are interested in the spatial part of the test particle acceleration. We have split the test particle acceleration in temporal and spatial part with respect to the observer congruence. The temporal projection along $u^\alpha$ leads to the evolution equation for the observed energy of the test particle along its world line; instead the spatial projection orthogonal to $u^\alpha$ leads to the evolution equation for the three-momentum of the test particle along its world line, where the kinematical quantities linked to the observer motion figure in these equations as \emph{inertial forces}.

The first term in Eq. (\ref{eq:S+T}) is called the \emph{spatial gravitational force}, interpreted as the inertial forces due to the motion of the observers themselves. These inertial forces involve the kinematical quantities of the observer congruence. Indeed, this term can be decomposed in \citep[see Sect. 9 in][]{Bini1997a,Bini1998}:
\begin{equation} \label{eq:GE+GM}
\gamma^2\frac{D_{\rm (tem)}(U,u)}{d\tau_{(U,u)}}u^\alpha=\gamma^2\left[a(u)^\alpha+H_{\rm (tem)}(u)_\beta^\alpha\,\nu(U,u)^\beta\right],
\end{equation}
where 
\begin{equation}
H_{\rm (tem)}(u)_\beta^\alpha=\begin{cases}
\quad\theta(u)^\alpha_\beta-\omega(u)^\alpha_\beta,& _{\rm tem=fw};\\
\quad2\theta(u)^\alpha_\beta-2\omega(u)^\alpha_\beta,& _{\rm tem=Lie};\\
\quad\theta(u)^\alpha_\beta-2\omega(u)^\alpha_\beta,& _{\rm tem=cfw}.
\end{cases}
\end{equation}
The first term in Eq. (\ref{eq:GE+GM}) leads to the \emph{gravitoelectric gravitational force}, instead the second term in Eq. (\ref{eq:GE+GM}) to the \emph{gravitomagnetic} or \emph{Coriolis gravitational force}. 

The second term in Eq. (\ref{eq:S+T}) can be decomposed into a longitudinal and transverse relative acceleration terms, with respect to the observer congruence, as \citep{Bini1997a,Bini1998,Bini2010}:
\begin{equation} \label{eq:CL+SC+TC}
\begin{aligned}
&\gamma\frac{D_{\rm (tem)}(U,u)}{d\tau_{(U,u)}}p(u,U)^\alpha\\
&=\gamma\frac{D_{\rm (tem)}(U,u)\,p(U,u)}{d\tau_{(U,u)}}\hat{\nu}(U,u)^\alpha\\
&+\gamma\,p(U,u)\frac{D_{\rm (tem)}(U,u)\,\hat{\nu}(U,u)^\alpha}{d\tau_{(U,u)}}\\
&=\frac{d p(U,u)^\alpha}{d\tau_U}+\gamma^2\, C_{\rm (tem)}(u)^\alpha_\beta\, \nu(U,u)^\beta\\
&\quad+\gamma^2\, \Gamma(u)^\alpha_{\beta\gamma}\,\nu(U,u)^\beta\,\nu(U,u)^\gamma,
\end{aligned}
\end{equation}
where we have divided the four-momentum, $p^\alpha=p\,\hat{\nu}^\alpha$, in the longitudinal part along the versor $\hat{\nu}(U,u)^\alpha$, called the \emph{relative centrifugal force}, and in the transverse part orthogonal to $\hat{\nu}(U,u)^\alpha$, called the \emph{relative centripetal force} \citep{Bini1997a,Bini1998}. The relative centripetal force can be written as:
\begin{equation} \label{eq:curvature}
\frac{D_{\rm (tem)}(U,u)\,\hat{\nu}(U,u)^\alpha}{d\tau_{(U,u)}}=k_{\rm (tem)}(U,u)\, \nu(U,u)^2\, \eta_{\rm (tem)}(U,u)^\alpha,
\end{equation}
where $\eta_{\rm (tem)}(U,u)^\alpha$ is the normal versor relative to the spatial test particle orbit in the osculating plane and $k_{\rm (tem)}(U,u)$ is the relative curvature. This term can be related to the curvature radius of the orbit, $\rho_{\rm (tem)}(U,u)$ through $k_{\rm (tem)}(U,u)=1/\rho_{\rm (tem)}(U,u)$ and also to the spatial connections $\Gamma(u)^\alpha_{\beta\gamma}$ through $\gamma^2\, \Gamma(u)^\alpha_{\beta\gamma}\,\nu(U,u)^\beta\,\nu(U,u)^\gamma$ \citep[][]{Bini1997a}. The evolution of the four-momentum explicitly reads as in the last row of Eq. (\ref{eq:CL+SC+TC}), where we have removed the contributions coming from the $\theta$-direction, since in our treatment the motion occurs only in the equatorial plane \citep[see][for the full description]{Bini1997a,Bini1997b,Bini1998,Bini1999}. In addition, the term $\Gamma(u)^\alpha_{\beta\gamma}\,\nu(U,u)^\beta\,\nu(U,u)^\gamma$ is called the \emph{space curvature force} in the threading point of view \citep{Bini1997a,Bini1998}.

\subsection{General relativistic PR EoMs in stationary and axially symmetric spacetimes}
We consider a stationary and axially symmetric spacetime, parametrized by the nonlinear reference frame associated to the Boyer-Lindquist coordinates $X^\alpha\equiv(t,r,\theta,\varphi)$. In such coordinates, the metric in the equatorial plane, $\theta=\pi/2$, reads as:
\begin{equation} \label{eq:metric}
ds^2=g_{tt}dt^2+g_{rr}dr^2+2g_{t\varphi}dt\, d\varphi+g_{\varphi\varphi}d\varphi^2,
\end{equation}
where all metric components depend only on $r$ and $\theta$. In such spacetimes, there are two kinds of observers: $(i)$ the hypersurface normal observers or zero angular momentum observers (ZAMOs) and $(ii)$ the threading observers following the time coordinate line trajectories. Both observers family are accelerated, because the firsts are dragged by the spinning central object, while the seconds are accelerating to contrast the frame dragging effect \citep{Bini1997a,Bini1997b,Bini1998}. In this environment, the ZAMO point of view is the easiest way to describe the motion of a test particle \citep[see Sect. 12.2 in][]{Bini1997a}. The ZAMO four-velocity is $u^\alpha=(N^{-1},0,-N^{-1}N^\varphi,0)$, where $N=(-g^{tt})^{-1/2}$ and $N^\varphi=g_{t\varphi}/g_{\varphi\varphi}$. The frame adapted to the ZAMOs is \citep{Bini2009,Bini2011}:
\begin{equation} \label{eq:ZAMO}
\begin{aligned}
&e^\alpha_t=u^\alpha,\ e^\alpha_r=\left(0,\frac{1}{\sqrt{g_{rr}}},0,0\right),\\
&\quad e^\alpha_\varphi=\left(0,0,\frac{1}{\sqrt{g_{\varphi\varphi}}},0,0\right).
\end{aligned}
\end{equation}
In the ZAMO point of view,  the metric, Eq. (\ref{eq:metric}), becomes \citep{Bini1997a,Bini2009,Bini2011}:
\begin{equation}
ds^2=-N^2dt^2+g_{\varphi\varphi}(d\varphi+N^\varphi dt)^2+g_{rr}dr^2.
\end{equation} 
For a stationary observer congruence, it is useful to exploit the intrinsic spatial Lie derivative, since it is the most appropriate to the spatial geometry without requiring additional kinematic linear transformations of the spatial tangent space, namely $C_{\rm (lie)}(u)^\alpha_\beta=0$, $\omega(u)^\alpha_\beta=0$, $H_{\rm (tem)}(u)^\alpha_\beta=2\theta(u)^\alpha_\beta$, $\Gamma(u)^r_{\varphi\varphi}=k_{\rm (Lie)}(u)^r$, and $\Gamma(u)^\varphi_{\varphi r}=-k_{\rm (Lie)}(u)^r$ \citep[see Sect 12.1 in][]{Bini1997a,Bini1998}. 

The test particle acceleration relative to the observer congruence, given by Eqs. (\ref{eq:S+T}), (\ref{eq:GE+GM}), and (\ref{eq:CL+SC+TC}), reads explicitly as \citep{Bini1997a,Bini1998,Bini2010}:
\begin{equation} \label{eq:GRacceleration}
\begin{aligned}
a(U)^\alpha&=\gamma^2\left[a(u)^\alpha+\Gamma(u)^\alpha_{\beta\gamma}\,\nu^\beta(U,u)\, \nu^\gamma(U,u)\right.\\
&\left.+2\theta(u)^\alpha_\beta\, \nu(U,u)^\beta\right]+\frac{d p(U,u)^\alpha}{d\tau(U,u)}\\
&=-F^{\rm (GE)}(U,u)^\alpha-F^{\rm (SC)}(U,u)^\alpha\\
&-F^{\rm (GM)}(U,u)^\alpha+\frac{d p(U,u)^\alpha}{d\tau_U},
\end{aligned}
\end{equation}
where the gravitational inertial forces are: gravitoelectric (GE), space curvature (SC), and gravitomagnetic (GM). This splitting, although it is very technical, permits to recognize and give an exact physical meaning to all terms contributing to characterize the acceleration $a(U)^\alpha$, as it happens for the classical case for Eq. (\ref{eq:acceleration}), see Sect. \ref{sec:classicPR} for further details. 

\begin{figure} [h!]
\includegraphics[trim=0cm 5.5cm 0cm 0cm,scale=0.45]{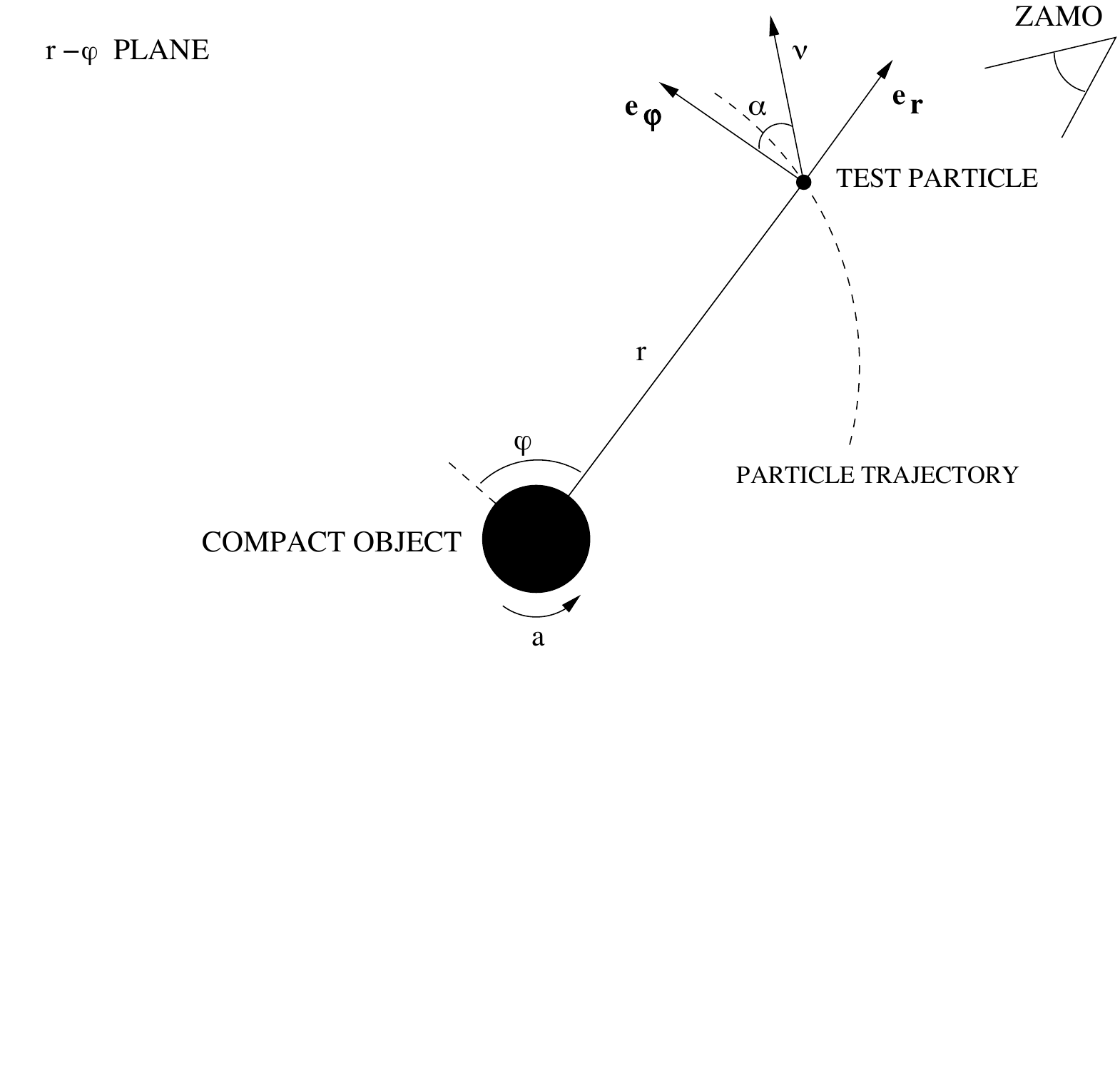}
\caption{The geometry of the problem is given by a test particle orbiting in the equatorial plane around a rotating compact object with spin $a$, at a radius $r$ and azimuthal angle $\varphi$. The test particle motion is described in the ZAMO reference frame $\{\mathbf{e_r},\mathbf{e_\varphi}\}$. The test particle moves with a velocity $\nu$, forming an angle $\alpha$ in the ZAMO reference frame.}
\label{fig:geometry}
\end{figure}

In presence of an external spatial force per unit test particle mass, $f(U)^\alpha$, the test particle EoMs are given by $a(U)^\alpha=f(U)^\alpha$. In our case, the external spatial force is represented by a radiation field, modeled as a coherent flux of photons moving along null geodesics, $k^\alpha$, on the background spacetime. The relative stress-energy tensor is \citep{Bini2009,Bini2011}:
\begin{equation}
T^{\alpha\beta}=\Phi^2 k^\alpha k^\beta, \quad k^\alpha k_\alpha=0,\quad k^\alpha\nabla_\alpha k=\beta=0,
\end{equation}
where $\Phi$ is the radiation intensity. We consider, that the photons can travel in any direction in the equatorial plane with angular momentum, $b\equiv L/E=-k_\varphi/k_t$. Therefore, it is useful to introduce the parameter $\beta$, defined as the azimuthal angle of the photon four-momentum measured in the local frame, $\{\mathbf{e_r},\mathbf{e_\varphi}\}$, related to the ZAMO \citep[see][for more details]{Bini2011}:
\begin{equation} \label{eq:k_observer}
\begin{aligned}
&k^\alpha=E(u)[u^\alpha+\hat{\nu}(k,u)^\alpha],\\
&\hat{\nu}(k,u)^\alpha=e_r^\alpha\,\sin\beta+e_\varphi^\alpha\, \cos\beta.
\end{aligned}
\end{equation}
In addition, the photon four-momentum can be also decomposed with respect to the test particle velocity, $U^\alpha$, in the following way \citep{Bini2009,Bini2011}:
\begin{equation} \label{eq:k_particle}
k^\alpha=E(U)[U^\alpha+\hat{V}(k,U)^\alpha].
\end{equation}
In this model, the radiation field is given by \citep{Bini2009,Bini2011}: 
\begin{equation} \label{eq:radiation}
F_{\rm(rad)}(U)^\alpha=-\frac{\sigma}{m}\, P(U)^\alpha_\beta\, T^\beta_\mu\, U^\mu,
\end{equation}
where $\sigma$ and $m$ are cross section and mass of the test particle, respectively. As done for the test particle velocity, we decompose also the photon four-momentum relative to the observer congruence in order to get the relative radiation force, $F_{\rm(rad)}(U,u)^\alpha$ \citep[see][for further details]{Bini2009,Bini2011}. In such decomposition, the intensity of the radiation is associated to the parameter $A$, defined as the emitted luminosity from the central source as seen by an observer at infinity in units of Eddington luminosity \citep[][]{Bini2009,Bini2011}. The explicit expression of the parameters figuring in Eq. (\ref{eq:GRacceleration}) can be found in the papers of \citet[][see Eqs. (2.7)]{Bini2009,Bini2011}. In such context, the important parameters to determine the motion of the test particle are: $\nu$ and $\alpha$, the velocity and azimuthal angle of the test particle in the ZAMO frame, respectively; $r$ and $\varphi$, the radius and the azimuthal angle in Boyer-Lindquist coordinates (see Fig. \ref{fig:geometry}).
Therefore using Eqs. (\ref{eq:GRacceleration}), (\ref{eq:radiation}), (\ref{eq:U_split}) the test particle EoMs are \citep{Bini2011}:
\begin{eqnarray} 
&&\frac{d\nu}{d\tau_U}=-\frac{\sin\alpha}{\gamma}[a(u)^{r}+2\nu\cos\alpha\,\theta(u)^{r}_{\varphi}] \label{EoM1}\\
&&\qquad+\frac{A(1+bN^\varphi)}{N^2(g_{\theta\theta}g_{\varphi\varphi})^{1/2}|\sin\beta|}[\cos(\alpha-\beta)-\nu][1-\nu\cos(\alpha-\beta)],\notag\\
&&\frac{d\alpha}{d\tau_U}=-\frac{\gamma\cos\alpha}{\nu}[a(u)^{r}+2\nu\cos\alpha\theta(u)^{r}_{\varphi}+\nu^2k_{\rm (lie)}(u)^{r}]\label{EoM2}\\
&&\qquad+\frac{A}{\nu}\frac{(1+bN^\varphi)[1-\nu\cos(\alpha-\beta)]}{N^2(g_{\theta\theta}g_{\varphi\varphi})^{1/2}|\sin\beta|}\sin(\beta-\alpha); \notag\\
&& U^r\equiv\frac{dr}{d\tau_U}=\frac{\gamma \nu\sin\alpha}{\sqrt{g_{rr}}},\label{EoM3}\\
&& U^\varphi\equiv\frac{d\varphi}{d\tau_U}=\frac{\gamma\nu\cos\alpha}{\sqrt{g_{\varphi\varphi}}}-\frac{\gamma N^\varphi}{N} \label{EoM4},
\end{eqnarray}
where $\Delta=r^2-2Mr+a^2$, and $b=\sqrt{g_{\varphi\varphi}}\cos\beta/[N(1-\nu_{\rm(s)}\cos\beta)]$ is the photon angular momentum with $\nu_{\rm (s)}=-2aM/(r\sqrt{\Delta})$. The test particle velocity $U^\alpha$ is obtained using Eq. (\ref{eq:U_split}), where the spatial velocity, $\nu^\alpha$, is decomposed in the adapted frame $\left\{\boldsymbol{e_t},\boldsymbol{e_r},\boldsymbol{e_\varphi},\boldsymbol{e_\theta}\right\}$ in this way: $\nu^\mu=\nu\sin\alpha\ e_r^\mu+\nu\cos\alpha\ e_\varphi^\mu$ \citep[see Eq. (2.14) in][for the explicit form]{Bini2009}.

It is important to note that Eq. (\ref{EoM1}), linked to the time component, is obtained from the orthogonality between $U^\alpha$ and $a(U)^\alpha$, that gives $a(U)^t=\nu[a(U)^r\sin\alpha+a(U)^\varphi\cos\alpha]$ for the acceleration \citep[see Sect. \ref{sec:observer} and Eq. (2.27) in][]{Bini2009}, and the orthogonality between $U^\alpha$ and $\hat{V}(k,U)^\alpha$, that gives $\hat{V}(k,U)^t=\nu[\hat{V}(k,U)^r\sin\alpha+\hat{V}(k,U)^\varphi\cos\alpha]$ for the force \citep[see Eq. (\ref{eq:k_particle}) and Eq. (2.27) in][]{Bini2009}. This condition underlines that the motion of the test particle, occurring in the equatorial plane around the central compact object, is determined classically and general relativistically by only two equations. However, $a(U)^t$ is present in the EoMs because it permits to determine the expression of $d\nu/d\tau_U$, that substituted in $a(U)^r$ (or equivalently in $a(U)^\varphi$) permits to get $d\alpha/d\tau_U$ \citep[see Eq. (2.29) in][]{Bini2009}. The addition of $U^r$ and $U^\varphi$ allows to univocally determine the four parameters $(\nu,\alpha,r,\varphi)$, characterizing the test particle motion (see Sect. \ref{sec:GRtoC}, for a further discussion). Eqs. (\ref{EoM1}) -- (\ref{EoM2}) describe the dynamics, instead Eqs. (\ref{EoM3}) -- (\ref{EoM4}) relate the test particle velocity components with respect to the dynamical variables. 

\section{The Lagrangian formulation of the Poynting-Robertson effect}
\label{sec:GRPR} 
We determine the  Lagrangian function and the Reyleigh dissipative function, that gives the EoMs (\ref{EoM1})--(\ref{EoM4}). We show, how the general relativistic formulation reduces to the classical case in the \emph{weak field limit} (mass over radius of the compact object tends to zero, $M/r\to0$, and velocities much lower than the speed of light, $v/c\ll1$).

\subsection{The general relativistic Lagrangian}
\label{sec:GR_lagrangian}
The aim of this section is to find the Lagrangian function, $\mathcal{L}_{\rm GR}$, and the Reyleigh dissipative function, $\mathcal{F}_{\rm GR}$, such that the relative ELEs, Eqs. (\ref{eq:ELeq}) give the general relativistic PR EoMs (\ref{EoM1})--(\ref{EoM4}). It is important to underline that the Lagrangian used in classical mechanics is for discrete particles, each with a finite number of degrees of freedom; instead the one used in field theory is a Lagrangian density applied to continua and fields, which have an infinite number of degrees of freedom. In absence of radiation, i.e., $A=0$, the test particle motion becomes purely geodetic. Therefore, the Lagrangian function coincides with its kinetic part and it is straightforward determined \citep{Misner1973,Chandrasekhar1992,Stephani2003}:
\begin{equation} \label{eq:GR_lagrangian}
\mathcal{L}_{\rm GR}\equiv \mathbb{T}_{\rm GR}-\mathbb{V}_{\rm GR}=\frac{1}{2}g_{\alpha\beta}\,U^\alpha\, U^\beta+\frac{1}{2},
\end{equation}
where the test particle four velocity, $U^\alpha$, is expressed in the ZAMO reference frame, \citep[see Eq. (2.29) in][]{Bini2009}. We note that Eq. (\ref{eq:GR_lagrangian}) includes also the contribution from the gravitational field, contained in the metric components $g_{\alpha\beta}$ (see Eq. (\ref{eq:lagrangian}) for comparison). The additive factor $1/2$ permits to obtain the gravitational potential in the weak field limit (see Sect. \ref{sec:GRtoC}). It is remarkable to note that such Lagrangian formalism is very general, because it can also be applied to a test particle moving outside the equatorial plane, since the $\theta$ velocity component, $U^\theta$, would be not null. 

\subsection{The general relativistic PR Reyleigh dissipative function}
\label{sec:GR_potential}
When the radiation field is present, i.e., $A\neq0$, we need to determine the potential of the radiation force, $F_{\rm(rad)}(U)^\alpha$. Based on the same arguments of Sect. \ref{sec:classicPR}, we have to find the Reyleigh dissipative function, $\mathcal{F}_{\rm GR}$, such that $Q^\alpha=-\partial\mathcal{F}_{\rm GR}/\partial U^\alpha$. In order to calculate such potential we have to verify that the radiation differential form, whose components are the $F_{\rm(rad)}(U)^\alpha$, is exact, namely it admits a primitive. The domain, where the radiation field is defined, is all the equatorial plane minus the region occupied by the compact object, that seems to be not simply connected. However, this domain, transformed in polar coordinates, is a rectangle, defined by $\varphi\in[0,2\pi]$ and $r\in(2M,+\infty]$, see Sect. \ref{sec:classicPR}. Therefore, we have to take care to check if the radiation differential form is closed, i.e., $\partial F_{\rm(rad)}(U)_\alpha/\partial U_\lambda=\partial F_{\rm(rad)}(U)_\lambda/\partial U_\alpha$. Therefore, calculating the cross derivatives:
\begin{equation}
\begin{aligned}
\frac{\partial F_{\rm(rad)}(U)_\alpha}{\partial U_\lambda}&=T_{\alpha\lambda}+\delta_{\alpha\lambda} U_\beta T^{\beta\mu}U_\mu\\&+U_\lambda T_\alpha^\mu U_\mu+U_\lambda U_\beta T_\alpha^\beta,\\
\frac{\partial F_{\rm(rad)}(U)_\lambda}{\partial U_\alpha}&=T_{\lambda\alpha}+\delta_{\lambda\alpha} U_\beta T^{\beta\mu}U_\mu\\&+U_\alpha T_\lambda^\mu U_\mu+U_\alpha U_\beta T_\lambda^\beta.
\end{aligned}
\end{equation}
and then equating them, we have:
\begin{equation} \label{eq:der_inc1}
U_\lambda\, T_\alpha^\mu\, U_\mu+U_\lambda\, U_\beta\, T_\alpha^\beta=U_\alpha\, T_\lambda^\mu\, U_\mu+U_\alpha\, U_\beta\, T_\lambda^\beta.
\end{equation}
Decomposing $T_{\alpha\beta}$ with respect to $U^\alpha$ as in Eq. (\ref{eq:k_particle}), Eq. (\ref{eq:der_inc1}) becomes:
\begin{equation} \label{eq:der_inc2}
U_\lambda\, k_\alpha=U_\alpha\, k_\lambda\quad \Leftrightarrow\quad U_\lambda\, \hat{V}_\alpha=U_\alpha\, \hat{V}_\lambda.
\end{equation}
If we multiply by scalar product both members of Eq. (\ref{eq:der_inc2}) for $U_\alpha$, we obtain $\hat{V}_\lambda=0$. This is a really strong condition, because it means that the radiation differential form is closed if and only if the radiation field is vanishing. 

An alternative way to find the Reyleigh dissipative function, $\mathcal{F}_{\rm GR}$, is in finding an integrating factor, $\mu=\mu(U)$, such that the new radiation differential form with components $\mu(U)\, F_{\rm(rad)}(U)^\alpha$, results to be exact. Calculating thus the cross derivatives and equating them, we have:
\begin{equation} \label{eq:eq:der_inc2}
\begin{aligned}
&\left(E(U)\frac{\partial \mu}{\partial U_\lambda}-2\mu k_\lambda\right)U_\alpha\\
&-\left(E(U)\frac{\partial \mu}{\partial U_\alpha}-2\mu k_\alpha\right)U_\lambda=0.
\end{aligned}
\end{equation}
Eq. (\ref{eq:eq:der_inc2}) reduces to solve the PDEs:
\begin{equation}
\left(E(U)\frac{\partial \mu}{\partial U_\lambda}-2\mu k_\lambda\right)=0,
\end{equation}
whose general solution, using the separation of variables method, is given by:
\begin{equation}
\mu(U)=\exp\left\{\frac{2\sum_\lambda (k_\lambda U_\lambda)}{E(U)}-2\right\}.
\end{equation}
The additive factor $-2$ permits to reduce to unity the integrating factor in the weak field limit (see Sect. \ref{sec:GRtoC}). The force field $F_{\rm(rad)}(U)^\alpha$ can be written equivalently as $(\nabla_{U^\alpha}\mathcal{F}_{\rm GR})/\mu$, preserving thus the dynamics described by EoMs (\ref{EoM1})--(\ref{EoM4}). The Reyleigh dissipative function does not depend on the photon geodesic structure, englobed in the stress-energy tensor $T_{\alpha\beta}$, because it is only function of the test particle velocity field $U^\alpha$. Therefore, this procedure can be also applied to more general photon geodesic emission, like for example for photons moving outside of the equatorial plane with variable $\theta$. We do not determine explicitly the functional form of the the general relativistic dissipative function $\mathcal{F}_{\rm GR}$, because it requires further analysis that could be the subject of another paper.

\subsection{Weak field approximation}
\label{sec:GRtoC}
We show how the general relativistic formalism in the weak field limit reduces to the classical case presented in Sect. \ref{sec:classicPR}. At this aim, we consider the test particle velocity and null geodesics in the Schwarzschild metric \citep{Misner1973,Bini2011}:
\begin{eqnarray}
U^\alpha&=&\left(\frac{\gamma}{\sqrt{1-\frac{2M}{r}}},\gamma\nu\sin\alpha\sqrt{1-\frac{2M}{r}},\frac{\gamma\nu\cos\alpha}{r},0\right),\\
k_\alpha&=&-E_p\left\{1,\frac{\left[1-\frac{b^2}{r^2}\left(1-\frac{2M}{r}\right)\right]^{1/2}}{1-\frac{2M}{r}},b,0\right\},
\end{eqnarray}
where $E_p$ is the photon energy depending on $c$, and $b$ the photon impact parameter. We remind that in the weak field limit $r\to\infty$, $b\to0$, and $\nu/c\to 0$ (as well as $\gamma\to1$).

The general relativistic Lagrangian, given by Eq. (\ref{eq:GR_lagrangian}), reduces to the classical Lagrangian, given by Eq. (\ref{eq:lagrangian}):
\begin{equation}
\mathcal{L}_{\rm GR}\approx -\frac{1}{2}\left(1-\frac{2M}{r}\right)+\frac{\nu^2}{2}+\frac{1}{2}=\frac{\nu^2}{2}+\frac{M}{r}\equiv\mathcal{L}_{\rm C},
\end{equation}
where in polar coordinates we have $\dot{r}=\nu\sin\alpha$ and $r\dot{\varphi}=\nu\cos\alpha$. The time component of the metric plus the additive factor gives the gravitational potential, instead the other components give the kinetic energy.

The integrating factor, $\mu$, does not figure in the classical case, because it reduces to unity. Indeed we have:
\begin{equation}
2\frac{\sum_\alpha (k^\alpha U^\alpha)}{E(U)}-2\approx2\left(\frac{-1+\nu\sin\alpha}{1+\nu\sin\alpha}\right)-2\to0,
\end{equation}
where $E(U)=-k_\alpha U^\alpha$, see Eq. (\ref{eq:k_particle}).
The integrating factor depends on the energy of the system, that in the newtonian case reduces to zero.

The analysis of the equations $a(U)^\alpha=F_{\rm (rad)}(U)^\alpha$ in the weak field limit is very interesting, because it gives a better physical explanation of the involved terms and it is possible to stress the fundamental role played by the general relativistic effects. We have the following approximations in $a(U)^\alpha$ \citep[see Eqs. (2.28) in][]{Bini2009}: $a(u)^r\approx M/r^2$, namely the gravitoelectric force corresponds to the gravitational force field; $k_{\rm (Lie)}(u)^r\approx -1/r+M/r^2$, where the first term represents the classical curvature radius, instead the second term, that does not figure in Eq. (\ref{eqm1}), is responsabile for the perihelion shift \citep[see Appendix in][for more details]{Bini2009}; $\theta(u)^r_\varphi\approx0$, because the spacetime is flat, therefore there is no deformation of the geodesic flow. Approximating $F_{\rm (rad)}(U)^\alpha$ through linear terms in $(\dot{r},\dot{\varphi})$, we have: $F_{\rm (rad)}(U)^r\approx A(1-2\dot{r})/r^2$, $F_{\rm (rad)}(U)^\varphi\approx -A\dot{\varphi}/r$, and $F_{\rm (rad)}(U)^t\approx -A/r^2-A\nu/r^2$. Therefore we have that $a(U)^r=F_{\rm (rad)}(U)^r$ reduces to Eq. (\ref{eqm1}), instead $a(U)^\varphi=F_{\rm (rad)}(U)^\varphi$ reduces to Eq. (\ref{eqm2}), as we would have expected. We underline that in the general relativistic case we have adopted geometrical units, i.e., $c=G=1$, in order not to have missing terms and create confusion with respect to the classical case. It is remarkable to note that $a(U)^t=F_{\rm (rad)}(U)^t$ reduces to the following equation:
\begin{equation}
\frac{d}{dt}\left(\frac{\nu^2}{2}+\frac{A-M}{r}\right)=-\frac{A}{r^2}\nu,
\end{equation}
describing the balance of the energy, where the right member represents the dissipated energy. In absence of the PR effect, i.e., $A\nu/r^2=0$, or the radiation field, i.e., $A=0$, we have the conservation of the energy.
 
\section{Conclusions}
\label{sec:conclusions}
In this work, we have developed the Lagrangian formulation of the general relativistic PR effect. The main challenges, that such work addresses and solves, are: the inverse problem, where the EoMs are given by \citet{Bini2009,Bini2011}, connected to the general relativistic radiation field including the PR effect, that is a dissipative force. A priori such problem might also not admit a Lagrangian formulation, due to the presence of a dissipative function \citep{Minguzzi2015}. In addition, another critical complication is the geometrical environment, constituted by a general stationary and axially symmetric spacetime, where the general relativistic effects contribute to make issue more thorny. This formulation may constitute an useful approach, among other existing methods, to investigate the general relativistic radiation fields including the PR effect. 

The importance to provide a Lagrangian formulation relies not only on a better understanding of the underlying physics, but also on a deeper analysis of the geometrical aspects and a simpler mathematical derivation of the EoMs, see Sect. \ref{sec:intro} and \ref{sec:GR_lagrangian}. This approach permitted also to have more insight in the radiation force itself and specifically in the PR effect, where the adding of an integrating factor, depending exponentially on the relativistic energy of the system, allow to identify the Reyleigh dissipative function, see \ref{sec:GR_potential}. The aim of such work is to furnish a complementary point of view in the study of the general relativistic PR effect with respect to the actual relativity of observer splitting formalism, increasing the interest on that subject and on the latter approach. In addition, comparing the classical and general relativistic Lagrangian formulations it is possible to recognize more clearly how GR influences the classical description, implying also an undimmed interpretations of the entailed variables, see Sect. \ref{sec:GR_formalism} and \ref{sec:GRtoC}. 

The results found in this paper pave principally the way at two compelling theoretical projects. The first one is into direction of improving the elementary description of the radiation field, with the inclusion of the PR effect, more adherent to describe the physical world \citep[see e.g.,][for further details]{Vaidya1951a,Vaidya1951b,Vaidya1973,Lindquist1965,Vaidya1999,Bini2011}. Indeed, the Lagrangian approach permits to more easily derive the relative EoMs and investigate the relative dynamical systems. 

The second proposal is in the actual and highlighted scientific research line of the theoretical study of the gravitational waves. Indeed in the linearized theory of GR a localized source, that is losing energy, emits gravitational waves, because for the energy conservation it must counterbalance the energy carried off by the gravitational radiation \citep[also known as gravitational radiation damping,][]{Misner1973}. This statement has been successfully confirmed by observations of the energy loss from the first discovered binary pulsar system PSR~B1913+16 \citep[][]{Taylor1982} and the most recently observations from two merging BHs \citep[see e.g.,][]{Abott2016a,Abott2016b} and a binary NS inspiral \citep{Abbott2017}. There is a strong analogy between gravitational waves and PR effect, because both are dissipative effects in GR. The Lagrangian approach and the results presented in this paper might be a valuable instrument in terms of theoretical understanding and subsequent observational testability of the gravitational waves.

\section*{Acknowledgements}
VDF and EB thank the International Space Science Institute in Bern for the support. VDF and EB acknowledge Prof. Donato Bini for his explanations on the relativity of observer splitting formalism. VDF and EB are grateful to Prof. Antonio Romano for the useful comments and valuable discussions aimed to improve this paper. We thank Prof. Luigi Stella for having inspired this work and for all the useful discussions.

\bibliography{references}

\end{document}